\documentclass[aps,twocolumn,showpacs]{revtex4}%
\usepackage{amsfonts}
\usepackage{amsmath}
\usepackage{amssymb}
\usepackage{appendix}
\usepackage{graphicx}%
\providecommand{\U}[1]{\protect\rule{.1in}{.1in}}
\providecommand{\U}[1]{\protect\rule{.1in}{.1in}}

\begin{document}
\title{The Physical Origin of Schr\"odinger Equation}
\author{\ Xiao-Bo Yan}
\email{xiaoboyan@126.com}
\affiliation{School of Physics and Electronic Engineering, Northeast Petroleum University, Daqing 163318, China}
\date{\today}


\begin{abstract}
Quantum mechanics is one of the basic theories of modern physics. Here, the famous Schr\"odinger equation and the differential operators representing mechanical quantities in quantum mechanics are derived, just based on the principle that the translation invariance (symmetry) of a system in Hamiltonian mechanics should be preserved in quantum mechanics. Moreover, according to the form of the differential operators, the commutation relation in quantum mechanics between the generalized coordinate and the generalized momentum can be directly obtained.  We believe that the results in this paper are very useful for understanding the physical origin of quantum mechanics.
\end{abstract}
\maketitle

\section{Introduction}

After about one century of development, quantum mechanics has become a fundamental theory in physics that provides a description of the physical properties at the scale of atoms and subatomic particles. It is the foundation of all quantum physics including quantum chemistry, quantum field theory, quantum technology and quantum information science.
Nevertheless, as a fundamental equation in quantum mechanics, the Schr\"odinger equation is given as a hypothesis in any quantum mechanics textbook, as well as the differential operators representing physical quantities. Many people believe that these basic concepts in quantum mechanics cannot be derived. In fact, they are closely related to the translation invariance (symmetry) in Hamiltonian mechanics where the equations of motion of a system are given as
\begin{eqnarray}
\dot{q}=\frac{\partial\,H}{\partial\,p},\quad
\dot{p}=-\frac{\partial\,H}{\partial\,q}.
\end{eqnarray}
Here, $q$ and $p$ are respectively the generalized coordinate and generalized momentum of the system (for simplicity, we assume that the system has only one generalized coordinate $q$ and the corresponding generalized momentum $p$), and $H(q,p,t)$ is the Hamiltonian function of the system. In Hamiltonian mechanics, the conserved quantities in a system are closely related to the translation invariance of the system. For instance, the generalized momentum $p$ in Eq. (1) will be a constant if the Hamiltonian $H(q,p,t)$ is invariant under the translation transformation with respect to the generalized coordinate $q$.

In the 1920s, quantum mechanics was founded by Schr\"odinger \cite{Schrodinger1926}, Heisenberg \cite{Heisenberg1925}, Born \cite{Born1925,Born1925II} and others based on some hypotheses including the hypothesis of wave-particle duality proposed by de Broglie \cite{Broglie1923}. Wave-particle duality is the concept in quantum mechanics that all particles exhibit a wave nature and vice versa. It means that every particle or quantum entity may be described as either a particle or a wave, and that the classical concepts “particle” or “wave” cannot fully describe the behavior of quantum-scale objects. Actually, any matter will not disappear in space and time, will only be converted from one form into another. Both ``particle" and ``wave" are the forms of matter. In 1927, the hypothesis of de Broglie wave was first conformed in Davisson-Germer experiment \cite{Davisson1928} which was an experimental milestone in the development of quantum mechanics.

Since the establishment of quantum wave mechanics in 1926, many noteworthy attempts have been made to derive the Schr\"odinger equation from different principles \cite{EDWARD1966,Field2011,Hall2002,Chavanis2017,Pelce1996,Briggs2001,Wieser2016,Palenik2014,Field2004,Bernstein2005,Ward2006,Rosen1964,Briggs2007}. Such as, in Ref. \cite{EDWARD1966} the author derived the Schr\"odinger equation with the hypothesis that every particle of mass $m$ is subject to a Brownian motion with a diffusion coefficient and no friction. In Ref. \cite{Field2011}, the Schr\"odinger equation is derived from the dynamical postulate of Feynman’s path integral formulation of quantum mechanics and the Hamilton--Jacobi equation of classical mechanics. In Ref. \cite{Hall2002}, the author derived the Schr\"odinger equation based on the exact uncertainty principle. It should be pointed out that no matter what method is used to derive the Schr\"odinger equation, the hypothesis that introduces Planck's constant $\hbar$ into quantum mechanics is essential.

Here, we give a novel, concise and heuristic way to give the Schr\"odinger equation just based on wave-particle duality and the translational symmetries (space and time translation invariance) of the system. We will see that it is the symmetry that builds a bridge from Hamiltonian mechanics to quantum mechanics. The material presented in this paper is of particular interest to university teachers of quantum mechanics courses and their students, at both undergraduate and graduate level. Researchers and educationalists on quantum mechanics, at all levels, may also benefit from a reading of the paper.

\section{Differential operator and eigen equation}

\textit{Although, in microscopic systems, the quantum entity behaves sometimes like a particle and sometimes like a wave, the symmetries of the system should be consistent no matter from the ``particle" or the ``wave" viewpoint.} 
According to this principle, the differential operators representing physical quantities and the Schr\"odinger equation in quantum mechanics can be easily obtained. In order to see that, let's assume that there is a particle whose classical behavior is dominated by the canonical equation in Eq. (1), and due to wave-particle duality, the corresponding wave can be denoted as a function $\psi(q,t)$ where $q$ is the generalized coordinate of the particle in Hamiltonian mechanics. (It can be conceivable that the wave function $\psi(q,t)$ must be related to the distribution of matter in the generalized coordinate space, since ``wave" is also a form of matter.) Now suppose that the particle system has a translation invariance (symmetry) with respect to the generalized coordinate $q$. As mentioned above, the corresponding wave $\psi(q,t)$ should have the same translation invariance, which means that the wave $\psi(q,t)$ and the shifted wave $\psi(q+a,t)$ should correspond to the same observables, here $a$ denotes the translational value of the generalized coordinate.

We will see that the translation invariance (symmetry) of the system will impose severe constraints on the wave function $\psi(q,t)$. First, the wave $\psi(q,t)$ cannot be a constant in the whole coordinate space because no interference fringes can be generated by such a wave. Secondly, the quantity $\psi^{*}(q,t)\psi(q,t)$ should be an observable (analogy with the observable light intensity in an electromagnetic field, i.e., $E^{*}(r,t)E(r,t)$, here $E(r,t)$ is just the electromagnetic wave at coordinate $r$ and time $t$) which will be invariant under the translation transformation, i.e., $\psi^{*}(q+a,t)\psi(q+a,t)=\psi^{*}(q,t)\psi(q,t)$. Hence, the only possible situation of the wave function $\psi(q,t)$ under the translation transformation is that
\begin{eqnarray}
\psi(q+a,t)=e^{iaf(q)}\psi(q,t)
\end{eqnarray}
with a real function $f(q)$.

In order to see what we can obtain from Eq. (2), we do the Taylor expansion on both sides of Eq. (2), i.e., 
\begin{eqnarray}
\psi(q+a,t)&=&\psi(q,t)+a\psi'(q,t)+\frac{a^{2}}{2}\psi''(q,t)+...\quad \\
e^{iaf(q)}\psi(q,t)&=&[1+iaf(q)-\frac{a^{2}f^{2}(q)}{2}+...]\psi(q,t).
\end{eqnarray}
Here, the prime denotes the derivative with respect to the generalized coordinate $q$.
Since Eq. (2) is true for any value $a$, the corresponding terms in Eqs. (3) and (4) must be equal. According to the second and third terms in Eqs. (3) and (4), we have
\begin{eqnarray}
\psi'(q,t)&=&if(q)\psi(q,t),\\
\psi''(q,t)&=&-f^{2}(q)\psi(q,t).
\end{eqnarray}   
Now we take the derivative of both sides of Eq. (5) with respect to $q$, we have
\begin{eqnarray}
\psi''(q,t)=if'(q)\psi(q,t)+if(q)\psi'(q,t).
\end{eqnarray}  
Substituting Eqs. (5) and (6) into Eq. (7) and using the fact that the wave $\psi(q,t)$ cannot be zero in the whole coordinate space, it is easy to obtain
\begin{eqnarray}
f'(q)=0,
\end{eqnarray}  
which means that the function $f(q)$ can only be a constant, denoted by $k$. Then according to Eq. (5), we have
\begin{eqnarray}
-i\frac{\partial}{\partial\,q}\psi(q,t)=k\psi(q,t),
\end{eqnarray} 
which is an eigen equation of the Hermitian operator $-i\frac{\partial}{\partial\,q}$ with the eigenfunction $\psi(q,t)$ and the eigenvalue $k$. The eigenvalue $k$ is a real number because in mathematics, the eigenvalues of any Hermitian operator are all real numbers.

\textit{What Eq. (9) means is that when a particle system has a translation invariance with respect to a generalized coordinate $q$ (the corresponding generalized momentum $p$ is a constant), the corresponding wave $\psi(q,t)$ must satisfy the eigen equation of the Hermitian operator $-i\frac{\partial}{\partial\,q}$, vice versa}. Hence, the eigenvalue $k$ in Eq. (9) should be directly related to the generalized momentum $p$. Meanwhile, the important Planck constant $\hbar$, which can reflect the quantum properties of a microscopic system, is obviously absent in Eq. (9). However, we can always multiply both sides of Eq. (9) by the Planck constant $\hbar$, then Eq. (9) becomes
\begin{eqnarray}
-i\hbar\frac{\partial}{\partial\,q}\psi(q,t)=\hbar k\psi(q,t).
\end{eqnarray}
In fact, the number $\hbar k$ in Eq. (10) is just the momentum $p$ of the particle, which can be directly verified by substituting the plane electromagnetic wave $e^{ikq}$ to $\psi(q,t)$ in Eq. (10) with the Einstein's hypothesis that the photon's momentum $p$ satisfies $p=\hbar k$ \cite{Einstein1905,Einstein1909,Einstein1917}. \textit{Therefore, it can be concluded that operating on the wave function $\psi(q,t)$, the Hermitian operator $-i\hbar\frac{\partial}{\partial\,q}$ will give out the value of momentum $p$ of the system, hence the Hermitian operator $-i\hbar\frac{\partial}{\partial\,q}$ can be considered as the generalized momentum operator (denoted by $\hat{p}$).} In fact, the momentum operator $-i\hbar\frac{\partial}{\partial\,q}$ is exactly the famous Born's assumption in the establishment of quantum mechanics, while here we know that the form of the momentum operator comes from the translation invariance (symmetry) in Hamiltonian mechanics.

If we take the complex conjugate to both sides of Eq. (10), we have
\begin{eqnarray}
i\hbar\frac{\partial}{\partial\,q}\psi^{*}(q,t)=p\psi^{*}(q,t)
\end{eqnarray} 
with $p=\hbar k$. 
It means that operating on the eigenfunction $\psi^{*}(q,t)$, the differential operator $i\hbar\frac{\partial}{\partial\,q}$ can also give out the same momentum $p$ of the system as the operator $-i\hbar\frac{\partial}{\partial\,q}$.  
Hence, we can also choose the differential operator $i\hbar\frac{\partial}{\partial\,q}$ as the generalized momentum operator $\hat{p}$ of the system.

\section{Commutation relation}

From the analysis above, we can see that due to wave-particle duality, the generalized momentum $p$ will be replaced with the momentum operator $\hat{p}$ $(=\pm i\hbar\frac{\partial}{\partial\,q})$. However, the generalized coordinate $q$ is still the same as that in Hamiltonian mechanics. The reason is that in the analysis above we study the wave function in generalized coordinate space (called coordinate representation). If we do that in generalized momentum space (called momentum representation), the situation is the opposite. It means that the generalized coordinate and momentum will be replaced by operators when the quantum entity exhibits the properties of waves, and in coordinate (momentum) representation, the coordinate (momentum) operator is the coordinate (momentum) itself.

Since the physical quantities (generalized coordinates and generalized momentum, as well as the Hamiltonian of the system) in quantum mechanics are represented by operators, the order in which they act on the wave function becomes important. The fundamental commutation relation (order relationship) between the generalized coordinate $q$ and the generalized momentum operator $\pm i\hbar\frac{\partial}{\partial\,q}$ can be directly obtained as
\begin{eqnarray}
[\,\pm i\hbar\frac{\partial}{\partial\,q},\,q\,]=\pm i\hbar,
\end{eqnarray} 
according to $[\,\frac{\partial}{\partial\,q},\,q\,]=1$ with the definition $[\hat{A},\,\hat{B}]=\hat{A}\hat{B}-\hat{B}\hat{A}$. According to Eq. (12), we can obtain the commutation relation between any operator and the Hamiltonian operator of the system.  
In fact, the equation with minus sign in Eq. (12) is exactly the Dirac's canonical quantization rule which we obtain very naturally here.

\section{Energy operator and Schr\"odinger equation}

In classical mechanics, the energy will be conserved if the Hamiltonian of the system is time independent. In this case, the system is said to have time translation invariance (symmetry). As previously mentioned, if there is some translation invariance in a particle system, then the corresponding wave must also have that.
It means that when the particle system has the time translation invariance, the corresponding wave function $\psi(q,t)$ must satisfy the condition that $\psi^{*}(q,t+t_{0})\psi(q,t+t_{0})=\psi^{*}(q,t)\psi(q,t)$, here $t_{0}$ denotes the translational value of time. Hence, the only possible situation of the shifted wave function $\psi(q,t+t_{0})$ is that
\begin{eqnarray}
\psi(q,t+t_{0})=e^{it_{0}f(t)}\psi(q,t)
\end{eqnarray}
with a real function $f(t)$.
Now we do the Taylor expansion on both sides of Eq. (13), i.e., 
\begin{eqnarray}
\psi(q,t+t_{0})&=&\psi(q,t)+t_{0}\psi'(q,t)+\frac{t_{0}^{2}}{2}\psi''(q,t)+...\quad\quad\\
e^{it_{0}f(t)}\psi(q,t)&=&[1+it_{0}f(t)-\frac{t_{0}^{2}f^{2}(t)}{2}+...]\psi(q,t).
\end{eqnarray}
Here, the prime denotes the derivative with respect to time $t$. Since Eq. (13) is true for any value $t_{0}$, the corresponding terms in Eqs. (14) and (15) must be equal. According to the second and third terms in Eqs. (14) and (15), we have
\begin{eqnarray}
\psi'(q,t)&=&if(t)\psi(q,t),\\
\psi''(q,t)&=&-f^{2}(t)\psi(q,t).
\end{eqnarray}   
According to Eqs. (16) and (17) and with the similar calculations in Eqs. (5)--(8), it is easy to obtain
\begin{eqnarray}
f'(t)=0.
\end{eqnarray}
It means that $f(t)$ can only be a constant. Hence, multipling both sides of Eq. (16) by the Planck constant $\hbar$, it is easy to obtain that the change of the wave function $\psi(q,t)$ with time will satisfy
\begin{eqnarray}
-i\hbar\frac{\partial}{\partial\,t}\psi(q,t)=E\psi(q,t).
\end{eqnarray}
Here, the constant $E$ should be the energy of the system, which can be verified by substituting the plane electromagnetic wave $e^{i\omega t}$ to $\psi(q,t)$ in Eq. (19) with the Einstein's hypothesis that the photon's energy $E$ satisfies $E=\hbar \omega$ \cite{Einstein1905,Einstein1909,Einstein1917}. 
\textit{Therefore, it can be concluded that operating on the wave function $\psi(q,t)$, the Hermitian operator $-i\hbar\frac{\partial}{\partial\,t}$ will give out the value of energy $E$ of the system.
Hence, the Hermitian operator $-i\hbar\frac{\partial}{\partial\,t}$ can be considered as the energy operator of the system.}
If we take the complex conjugate to both sides of Eq. (19), we have
\begin{eqnarray}
i\hbar\frac{\partial}{\partial\,t}\psi^{*}(q,t)=E\psi^{*}(q,t).
\end{eqnarray}
It means that operating on the eigenfunction $\psi^{*}(q,t)$, the differential operator $i\hbar\frac{\partial}{\partial\,t}$ can give the same energy $E$ of the system.
Hence, we can also choose $i\hbar\frac{\partial}{\partial\,t}$ as the energy operator of the system.

Meanwhile, there exists another energy operator in the system, i.e., the Hamiltonian operator $\hat{H}(\hat{q},\hat{p},t)$ since the Hamiltonian function $H(q,p,t)$ is just the energy of the system in Hamiltonian mechanics. \textit{It means that the two Hermitian operators are equivalent when they operate on a wave function, i.e.,}
\begin{eqnarray}
\pm i\hbar\frac{\partial}{\partial\,t}\equiv\hat{H}(\hat{q},\hat{p},t).
\end{eqnarray} 
Hence, according to Eq. (21), the time evolution of a wave function $\psi(q,t)$ will satisfy
\begin{eqnarray}
\pm i\hbar\frac{\partial}{\partial\,t}\psi(q,t)=\hat{H}(\hat{q},\hat{p},t)\psi(q,t).
\end{eqnarray} 
In fact, if we choose one equation in Eq. (22) (for example, the one with plus sign) to describe the behavior of a particle, then the other equation will describe the behavior of an antiparticle. 
It can be seen from Eq. (22) that if we choose $-i\hbar\frac{\partial}{\partial\,q}$ as the momentum operator $\hat{p}$, the equation with plus sign in Eq. (22) is just the famous Schr\"odinger equation in quantum mechanics.

\section{Conclusion}

In summary, due to wave-particle duality in microscopic systems, Hamiltonian mechanics describing the macroscopic properties of a system has to be replaced by quantum mechanics when we study the properties of quantum-scale objects.
According to the principle that the symmetries of a system must be consistent in quantum mechanics and Hamiltonian mechanics, we obtain the famous Schr\"odinger equation and the form of the differential operators representing mechanical quantities in quantum mechanics. The fact that the wave function satisfies the eigen equation of some differential operator is exactly the reflection of the symmetry of the system. In addition, according to the form of the differential operator, the commutation relation between the generalized coordinate operator and the generalized momentum operator can be directly obtained.


\bigskip

\begin{thebibliography}{99}                                                                                               %


\bibitem {Schrodinger1926} E. Schr\"odinger, ``An Undulatory Theory of the Mechanics of Atoms and Molecules", Physical Review \textbf{28}, 1049 (1926).


\bibitem {Heisenberg1925} W. Heisenberg, ``\"Uber quantentheoretische Umdeutung kinematischer und mechanischer Beziehungen", Zeitschrift f\"ur Physik (in German), \textbf{33}, 879--893 (1925).


\bibitem{Born1925} M. Born and P. Jordan, ``Zur Quantenmechanik", Zeitschrift f\"ur Physik, \textbf{34}, 858--888 (1925).


\bibitem{Born1925II} M. Born, W. Heisenberg, and P. Jordan, ``Zur Quantenmechanik II", Zeitschrift f\"ur Physik, \textbf{35}, 557--615 (1926).


\bibitem {Broglie1923} L. de Broglie, ``Waves and quanta", Nature \textbf{112}, 540 (1923).


\bibitem {Davisson1928} C. J. Davisson and L. H. Germer, ``Reflection of Electrons by a Crystal of Nickel", \textit{Proceedings of the National Academy of Sciences of the United States of America} \textbf{14}, 317--322 (1928).


\bibitem{EDWARD1966} E. Nelson, ``Derivation of the Schr\"odinger Equation from Newtonian Mechanics", Phys. Rev.
\textbf{150}, 1079--1085 (1966).

\bibitem{Field2011} J. H. Field, ``Derivation of the Schr\"odinger equation from the Hamilton--Jacobi equation in Feynman’s path integral formulation of quantum mechanics", Eur. J. Phys. \textbf{32}, 63–87 (2011).


\bibitem{Hall2002} M. J. W. Hall and M. Reginatto, ``Schr\"odinger equation from an exact uncertainty
principle", J. Phys. A \textbf{35}, 3289--3303 (2002).


\bibitem{Chavanis2017}P. H. Chavanis, ``Derivation of a generalized Schr\"odinger equation from the theory
of scale relativity", Eur. Phys. J. Plus \textbf{132}, 286 (2017). 


\bibitem{Pelce1996}P. Pelce, ``Another derivation of the
Schr\"odinger equation", Eur. J. Phys. \textbf{17}, 116--117 (1996).



\bibitem{Briggs2001} J. S. Briggs and J. M. Rost, ``On the derivation of the time--dependent
equation of Schr\"odinger", Foundations of Physics \textbf{31}, 693--712 (2001).

\bibitem{Wieser2016} R. Wieser, ``Derivation of a time dependent Schr\"odinger equation as the quantum mechanical Landau--Lifshitz--Bloch equation", J. Phys.: Condens. Matter \textbf{28}, 396003 (2016).


\bibitem{Palenik2014} M. C. Palenik, ``Quantum mechanics from Newton’s second law and the canonical commutation
relation $[X,P]=i$", Eur. J. Phys. \textbf{35}, 045014 (2014).

\bibitem{Field2004}J. H. Field, ``Relationship of quantum mechanics to classical electromagnetism and classical relativistic mechanics", Eur. J. Phys. \textbf{25}, 385 (2004).


\bibitem{Bernstein2005} J. Bernstein, ``Max Born and the quantum theory", Am. J. Phys. \textbf{73}, 999 (2005).

\bibitem{Ward2006} D. W. Ward and S. Volkmer, ``How to Derive the Schr\"odinger Equation",
arXiv:physics/0610121v1.

\bibitem{Rosen1964} N. Rosen, ``The relation between classical and quantum mechanics", Am. J. Phys. \textbf{32}, 597 (1964).

\bibitem{Briggs2007}J. S. Briggs, S. Boonchui, and S. Khemmani, ``The derivation of time-dependent schr\"odinger
equations", Journal of Physics A: Mathematical and Theoretical, \textbf{40}, 1289--1302 (2007).



\bibitem{Einstein1905} A. Einstein, ``\"Uber einen die Erzeugung und Verwandlung des Lichtes betreffenden heuristischen Gesichtspunkt", Annalen der Physik (in German), \textbf{17}, 132--148 (1905).


\bibitem{Einstein1909} A. Einstein, ``\"Uber die Entwicklung unserer Anschauungen \"uber das Wesen und die Konstitution der Strahlung", Physikalische Zeitschrift (in German), \textbf{10}, 817--825 (1909).


\bibitem{Einstein1917} A. Einstein, ``Zur Quantentheorie der Strahlung", Physikalische Zeitschrift (in German), \textbf{18}, 121--128 (1917).









\end{thebibliography}
\end{document}